\documentclass[12pt]{iopart}

\usepackage{iopams}  
\usepackage{graphicx}

\usepackage{rsfs}    

\newcommand{\bra}{\langle}
\newcommand{\ket}{\rangle}
\newcommand{\vect}[1]{\mbox{\boldmath $ #1 $}}

\newcommand{\Trace}{\mbox{Tr}}

\begin{document}

\title[Photon detection operator and complementarity]{%
Photon detection operator and
complementarity between electric detector and magnetic detector}

\author{Shogo Tanimura}

\address{Department of Complex Systems Science,
Graduate School of Information Science

Nagoya University,
Nagoya 464-8601, Japan}
\ead{tanimura@is.nagoya-u.ac.jp}
\begin{abstract}
It had been a long standing problem that there is no consistent definition
of photon position operator nor photon number density 
in the context of quantum theory.
In this paper we derive
the photon detection operator,
which defines location of photon absorption,
by applying the theory of indirect measurement
to quantum electrodynamics.
It is shown that
the photon detection probability depends on
properties of a photon-absorbing atom,
in particular, 
on both electric and magnetic dipole moments of the atom.
An experiment is proposed,
in which the complementarity of wave-particle nature of light will be tested.
It is also discussed that
the complementarity is related 
to the non-commutativity of the electric and the magnetic fields.\footnote{%
This is a manuscript for a proceedings of the conference,
the 20th Central European Workshop on Quantum Optics (CEWQO2013)
held at the Royal Institute of Technology (KTH) in Stockholm, Sweden
during 16-20 June, 2013.
When this work was presented at the conference, 
it had a title
``Photon position operators and
complementarity between electric detector and magnetic detector.''}
\end{abstract}

\vspace{2pc}
\noindent{\it Keywords}: quantum optics, photon, position, complementarity,
measurement model

\maketitle

\section{Introduction}
The statistical interpretation, which was proposed by Born,
is at the heart of quantum theory.
For a wave function 
$ \psi ( x,y,z,t ) $
of an electron, the square of its absolute value
\begin{equation}
	| \psi ( x,y,z,t ) |^2 \, \Delta x \, \Delta y \, \Delta z
	\label{Born}
\end{equation}
is proportional to a probability
for finding the electron in a volume 
$ \Delta x \, \Delta y \, \Delta z $
around the point $ ( x,y,z ) $.
The probability satisfies the local conservation law 
\begin{equation}
	\rho = | \psi |^2,
	\qquad
	\vect{j} = - \frac{i \hbar}{2m}
	( \psi^* \nabla \psi - \nabla \psi^* \psi ),
	\qquad
	\frac{\partial \rho}{\partial t} + \mbox{div} \, \vect{j} = 0.
	\label{local conservation}
\end{equation}
The position and the momentum of the electron 
are represented by operators
$ \hat{x}_j $ and $ \hat{p}_j $ $ (j=1,2,3) $ respectively.
They act on the wave function as
\begin{equation}
	\hat{x}_j \psi = x_j \psi,
	\qquad
	\hat{p}_j \psi = - i \hbar \frac{\partial}{\partial x_j} \psi,
\end{equation}
and they satisfy the canonical commutation relation
\begin{equation}
	[ \hat{x}_j, \hat{x}_k ] = 0,
	\qquad
	[ \hat{p}_j, \hat{p}_k ] = 0,
	\qquad
	[ \hat{x}_j, \hat{p}_k ] = i \hbar \delta_{jk}.
	\label{CCR}
\end{equation}
Other massive particles like protons, neutrons, and atoms
can be described in a similar way.

However, the above standard scheme is not applicable to photons,
which are massless spin 1 particles.
Pauli~\cite{Pauli:1933} noted that
in quantum field theory
there does not exist photon number density 
satisfying the local conservation law (\ref{local conservation}).
Pryce~\cite{Pryce:1948} showed that 
it is impossible to implement
photon position operators satisfying the CCR (\ref{CCR}).
Newton, Wigner~\cite{Newton:1949} and Wightman~\cite{Wightman:1962}
proved that there is no localized state (position eigenstate) of photon.
Thus, 
geometric notions like position and number density of photons
cannot be defined in a naive manner.

Other researchers
have defined
localized states, 
position operators, 
and probability density for photons
in more elaborated manners.
Bialynicki-Birula \cite{Bialynicki-Birula:1998}
constructed one-photon state whose
energy density is localized with exponential falloff,
although it does not represent a stationary state.
Hawton \cite{Hawton:1999} constructed
photon position operators that satisfy $ [ \hat{x}_j, \hat{x}_k ] = 0 $.
Keller \cite{Keller:2000} provided 
a wave function formalism for describing
photon emission process in the space-time.

In this paper we formulate photon detection operator
which characterizes probability and space-time location 
of photon emission or absorption.
We take physical properties of photon-detecting atoms 
into our formalism explicitly.
Our formulation can describe finite-time processes
and also photon-detection  processes that involve magnetic field-matter couplings.
As an application of our formalism
we propose a scheme of experiment to test wave-particle complementarity of light.

\section{Indirect measurement model}
Here we describe a general scheme of indirect measurement model.
An object system has a Hilbert space $ \mathscr{H} $
and a measuring apparatus has a Hilbert space $ \mathscr{K} $.
Initial states of the object system and of the apparatus are characterized with
density matrices $ \hat{\varrho} $ and $ \hat{\sigma} $, respectively.
Interaction between the object and the apparatus
is described by a unitary 
operator $ \hat{U} $ acting on 
$ \mathscr{H} \otimes \mathscr{K} $.
The apparatus has a self-adjoint operator $ \hat{M} $ 
which plays a role of a meter observable.
The operator $ \hat{M} $ admits a spectral decomposition
\begin{equation}
	\hat{M} = \sum_r m_r \hat{P}_r,
	\label{meter}
\end{equation}
where $ \{ \hat{P}_r \} $ are projection-valued measure satisfying
\begin{equation}
	\hat{P}_r^\dagger =  \hat{P}_r,
	\qquad
	\hat{P}_r \hat{P}_s = \delta_{rs} \hat{P}_r,
	\qquad
	\sum_r \hat{P}_r = 1.
\end{equation}
After the interaction process we read out the meter.
The probability for reading the value $ m_r $ as the meter output 
is calculated with the Born statistical formula
\begin{equation}
	p_r =
	\Trace_{ \mathscr{H} \otimes \mathscr{K} }
	( \hat{P}_r \, U \hat{\varrho} \otimes \hat{\sigma} U^\dagger )
	\label{probability}
\end{equation}
and the density matrix of the object system after the measurement is given by
\begin{equation}
	T_r ( \hat{\varrho} )
	=
	\frac{1}{ \: p_r \: }
	\Trace_{ \mathscr{K} }
	( \hat{P}_r \, U \hat{\varrho} \otimes \hat{\sigma} U^\dagger ).
	\label{output state}
\end{equation}

\section{Photon detection operator}
Here we shall apply the above scheme
to photon detection process.
Photons are regarded as an object system
and the apparatus consists of electrons in atoms.
We describe dynamics of the whole system in the interaction picture
and use the Coulomb gauge, 
in which the vector potential $ \vect{A} $ satisfies
$ \mbox{div} \, \vect{A} = 0 $.
The free photons and atoms obey the Hamiltonian 
\begin{equation}
	\hat{H}_0
	= 
	\frac{1}{2} ( \vect{\hat{E}}^2 + \vect{\hat{B}}^2 )
	+ \hat{H}_{\mbox{\tiny atom}}.
	\label{free H}
\end{equation}
The interaction between photons and atoms 
is described by the minimal coupling 
\begin{equation}
	\hat{H}_{\mbox{\tiny int}}
	= - 
	\int \vect{\hat{A}} ( \vect{x}, t ) 
	\cdot \vect{\hat{J}} ( \vect{x}, t ) \, d^3 x,
	\label{interaction H}
\end{equation}
where $ \vect{\hat{J}} $ is the electric current operator of the electrons.
Then the time-evolution unitary operator is given by
\begin{equation}
	\hat{U} 
	= 
	{\cal T} 
	\exp \bigg[ \!
	- \frac{i}{\hbar} 
	\int_{t_0}^{t_1} \! \hat{H}_{\mbox{\tiny int}} \, dt
	\bigg]
	= 
	{\cal T} 
	\sum_{n=0}^\infty
	\frac{1}{\: n! \:}
	\bigg( \!
	- \frac{i}{\hbar} 
	\int_{t_0}^{t_1} \! \hat{H}_{\mbox{\tiny int}} \, dt
	\bigg)^n,
	\label{interaction U}
\end{equation}
where the symbol $ {\cal T} $ denotes the time-ordered products of operators.
By substituting (\ref{interaction U}) into (\ref{probability})
and by taking the first-order term in the expansion (\ref{interaction U}),
we obtain an approximate probability of single-photon absorption
\begin{equation}
	p_r 
	\, \sim \,
	\frac{1}{\hbar^2}
	\Trace_{ \mathscr{H} \otimes \mathscr{K} }
	\Bigg( \hat{P}_r 
	\int_{t_0}^{t_1} \! \hat{H}_{\mbox{\tiny int}} dt
	\, \hat{\varrho} \otimes \hat{\sigma} 
	\int_{t_0}^{t_1} \! \hat{H}_{\mbox{\tiny int}} dt
	\Bigg).
	\label{1st probability}
\end{equation}
Moreover, 
the initial state of the apparatus $ \hat{\sigma} $ is assumed to be 
the ground state of the Hamiltonian 
$ \hat{H}_{\mbox{\tiny atom}} $ as
\begin{equation}
	\hat{\sigma} =
	| \epsilon_0 \ket \bra \epsilon_0 |,
	\qquad
	\hat{H}_{\mbox{\tiny atom}} | \epsilon_0 \ket =
	\epsilon_0 | \epsilon_0 \ket,
	\label{initial state}
\end{equation}
and the final state $ \hat{P}_r $ of the apparatus is assumed to be 
an excited state
\begin{equation}
	\hat{P}_r =
	| \epsilon_r \ket \bra \epsilon_r |,
	\qquad
	\hat{H}_{\mbox{\tiny atom}} | \epsilon_r \ket =
	\epsilon_r | \epsilon_r \ket.
	\label{final state}
\end{equation}
In the interaction picture, the time-dependent operator is defined by
\begin{equation}
	\hat{H}_{\mbox{\tiny int}} (t)
	=
	e^{i \hat{H}_0 t/ \hbar}
	\, \hat{H}_{\mbox{\tiny int}} \,
	e^{-i \hat{H}_0 t/ \hbar}.
	\label{interaction picture}
\end{equation}
Thus, the detection probability (\ref{1st probability}) becomes
\begin{equation}
	p_r 
	\, \sim \,
	\frac{1}{\hbar^2}
	\Trace_{ \mathscr{H} }
	\Bigg(
	\int_{t_0}^{t_1} \! \vect{\hat{A}}
	\cdot
	\bra \epsilon_r | \vect{\hat{J}} | \epsilon_0 \ket
	d^3 x dt
	\; \hat{\varrho} 
	\int_{t_0}^{t_1} \! \vect{\hat{A}}
	\cdot
	\bra \epsilon_0 | \vect{\hat{J}} | \epsilon_r \ket
	d^3 x dt
	\Bigg).
	\label{2nd probability}
\end{equation}
By introducing the detection operator
\begin{eqnarray}
	\hat{D}_r 
&=&
	\frac{1}{\hbar}
	\int \!\!\!
	\int_{t_0}^{t_1} \! 
	\vect{\hat{A}} ( \vect{x}, t )
	\cdot
	\bra \epsilon_r | \vect{\hat{J}} ( \vect{x}, t ) | \epsilon_0 \ket
	d^3 x \, dt
	\nonumber \\
&=&
	\frac{1}{\hbar}
	\int \!\!\!
	\int_{t_0}^{t_1} \! 
	\vect{\hat{A}} ( \vect{x}, t )
	\cdot
	\bra \epsilon_r | \vect{\hat{J}} ( \vect{x}, 0 ) | \epsilon_0 \ket
	\, e^{i ( \epsilon_r - \epsilon_0 ) t / \hbar } \,
	d^3 x \, dt,
	\label{detection operator}
\end{eqnarray}
we can write the photon detection probability and the state after the detection as
\begin{equation}
	p_r 
	\sim 
	\Trace_{ \mathscr{H} }
	( \hat{D}_r^\dagger \hat{D}_r 
	\, \hat{\varrho} ),
	\qquad
	T_r ( \hat{\varrho} )
	=
	\hat{D}_r \, \hat{\varrho} \, \hat{D}_r^\dagger.
	\label{result}
\end{equation}
In the Coulomb gauge, the vector potential is expanded in plane waves as
\begin{eqnarray}
	\vect{\hat{A}} ( \vect{x}, t )
&=&
	\int \frac{\sqrt{\hbar} \, d^3 k}{ \sqrt{ 2 \omega_k (2 \pi)^3 }}
	\sum_{s = 1,2}
	\Big(
	\vect{\varepsilon}_{ks} \hat{a}_{ks} 
	\, e^{i ( {\bf k} \cdot {\bf x} - \omega t) }
	+ 
	\vect{\varepsilon}_{ks}^* \hat{a}_{ks}^\dagger
	\, e^{-i ( {\bf k} \cdot {\bf x} - \omega t) }
	\Big)
	\nonumber \\
&=&
	\vect{\hat{A}}^{(+)} ( \vect{x}, t ) +
	\vect{\hat{A}}^{(-)} ( \vect{x}, t )
	\label{Fourier}
\end{eqnarray}
with the frequency $ \omega_k = c | \vect{k} | $ 
and the transverse polarization vectors $ \vect{\varepsilon}_{ks} $
satisfying $ \vect{k} \cdot \vect{\varepsilon}_{ks} = 0 $.
The first term including the photon annihilation operators $ \hat{a}_{ks} $
is called the positive frequency part of the electromagnetic field
while 
the second term including the creation operators $ \hat{a}_{ks}^\dagger $
is called the negative frequency part.
The Fourier transform of the matrix element of the electric current operator 
are denoted as
\begin{equation}
	\vect{J}_{kr}
	=
	\int d^3 x
	\bra \epsilon_r | \vect{\hat{J}} ( \vect{x}, 0 ) | \epsilon_0 \ket
	\, e^{i {\bf k} \cdot {\bf x} }.
\end{equation}
Then the detection operator is rewritten as
\begin{eqnarray}
	\hat{D}_r 
&=&
	\int \! \frac{d^3 k}{ \sqrt{ 2 \hbar \omega_k (2 \pi)^3 }}
	\sum_{s = 1,2}
	\nonumber \\
	&& \quad
	\int_{t_0}^{t_1} \! dt \,
	(
	\vect{\varepsilon}_{ks} \cdot \vect{J}_{kr} 
	\, \hat{a}_{ks} \, e^{ - i \omega t }
	+ 
	\vect{\varepsilon}_{ks}^* \cdot \vect{J}_{-kr} 
	\, \hat{a}_{ks}^\dagger \, e^{ i \omega t) }
	) 
	\, e^{i ( \epsilon_r - \epsilon_0 ) t / \hbar }.
	\label{Fourier detection operator}
\end{eqnarray}

If we take the limit $ t_0 \to - \infty $ and $ t_1 \to \infty $,
the time integrals in the above equation become
\begin{eqnarray}
&&	\int_{t_0}^{t_1} \! dt \,
	 e^{ - i \omega t }
	\, e^{i ( \epsilon_r - \epsilon_0 ) t / \hbar }
	\: \to \:
	2 \pi \hbar \, \delta ( \epsilon_r - \epsilon_0 - \hbar \omega ),
	\label{time integral 1} \\
&&	\int_{t_0}^{t_1} \! dt \,
	 e^{ i \omega t }
	\, e^{i ( \epsilon_r - \epsilon_0 ) t / \hbar }
	\: \to \:
	2 \pi \hbar \, \delta ( \epsilon_r - \epsilon_0 + \hbar \omega ).
	\label{time integral 2}
\end{eqnarray}
When an atom excitation process with $ \epsilon_r - \epsilon_0 > 0 $ is concerned,
the second integral (\ref{time integral 2}) vanishes.
Therefore, it is justified to remove the negative frequency part 
and to leave only the positive frequency (photon absorption) part 
of the electromagnetic field
in (\ref{detection operator}) for a long-time measurement process.
Thus, the detection operator is reduced to 
\begin{eqnarray}
	\hat{D}_r^{(+)}
&=&
	\int \! \frac{ \sqrt{\hbar} \, d^3 k}{ \sqrt{ 2 \omega_k (2 \pi)^3 }}
	\sum_{s = 1,2}
	\vect{\varepsilon}_{ks} \cdot \vect{J}_{kr} 
	\, \hat{a}_{ks} \, 
	2 \pi \, \delta ( \epsilon_r - \epsilon_0 - \hbar \omega )
	\nonumber \\
&=&
	\frac{1}{\hbar}
	\int \!\!\!
	\int_{- \infty}^{\infty} \! 
	\vect{\hat{A}}^{(+)} ( \vect{x}, t )
	\cdot
	\bra \epsilon_r | \vect{\hat{J}} ( \vect{x}, 0 ) | \epsilon_0 \ket
	\, e^{i ( \epsilon_r - \epsilon_0 ) t / \hbar } \,
	d^3 x \, dt.
	\label{annihilation process}
\end{eqnarray}

However, for a short-time process,
the limits (\ref{time integral 1}), (\ref{time integral 2}) cannot be justified
and the original form (\ref{Fourier detection operator}) 
of the detection operator should be used.
In that case the relation 
$ \epsilon_r - \epsilon_0 = \hbar \omega $
does not exactly hold and a natural line width is observed.


\section{Complementarity}
An electrically neutral atom can interact with electromagnetic field
via electric or magnetic dipole moment couplings.
An electric polarization density $ \vect{d} $
and a magnetization density $ \vect{m} $ generate electric current
\begin{equation}
	\vect{\hat{J}} 
	=
	\frac{\partial \vect{\hat{d}}}{\partial t}
	+ \mbox{rot} \, \vect{\hat{m}}.
	\label{induced current}
\end{equation}
By substituting this into (\ref{interaction H}) and integrating by parts, 
we make the interaction Hamiltonian in the form
\begin{eqnarray}
	- 
	\int \vect{\hat{A}} \cdot \vect{\hat{J}} \, d^3 x \, dt
&=&
	- \!\! \int 
	\Big(
	- \frac{\partial \vect{\hat{A}}}{\partial t}
	\cdot \vect{\hat{d}}
	+ \mbox{rot} \, \vect{\hat{A}}
	\, \cdot \vect{\hat{m}}
	\Big) d^3 x \, dt
\nonumber \\
&=&
	- \!\! \int 
	(
	\vect{\hat{E}} \cdot \vect{\hat{d}}
	+ \vect{\hat{B}} \cdot \vect{\hat{m}}
	) d^3 x \, dt
	\label{H'}
\end{eqnarray}
and rewrite the detection operator (\ref{annihilation process}) in the form
\begin{eqnarray}
	\hat{D}_r^{(+)}
&=&
	\frac{1}{\hbar}
	\int 
	\Big(
	\vect{\hat{E}}^{(+)} \! 
	\cdot
	\bra \epsilon_r | \vect{\hat{d}} 
	| \epsilon_0 \ket
	+
	\vect{\hat{B}}^{(+)} \! 
	\cdot
	\bra \epsilon_r | \vect{\hat{m}} 
	| \epsilon_0 \ket
	\Big)
	d^3 x \, dt.
\end{eqnarray}
This expression justifies
Glauber's proposal~\cite{Glauber:1963}
for using the matrix element of 
the positive frequency part of the electric field
$ \bra \mbox{vac} | \vect{\hat{E}}^{(+)} ( \vect{x}, t ) 
| \mbox{photon} \ket $
as a probability amplitude for photon detection.
However, if the electric dipole moment 
$ \bra \epsilon_r | \vect{\hat{d}} | \epsilon_0 \ket $
of the detector atom is zero,
the magnetic dipole moment 
$ \bra \epsilon_r | \vect{\hat{m}} | \epsilon_0 \ket $
becomes relevant as the next leading term.
Thus, the magnetic field amplitude
$ \bra \mbox{vac} | \vect{\hat{B}}^{(+)} ( \vect{x}, t ) 
| \mbox{photon} \ket $
also should be taken into account for photon detection.

\begin{figure}[t]
\begin{center}
\vspace{-8mm}
\scalebox{0.55}{
\includegraphics{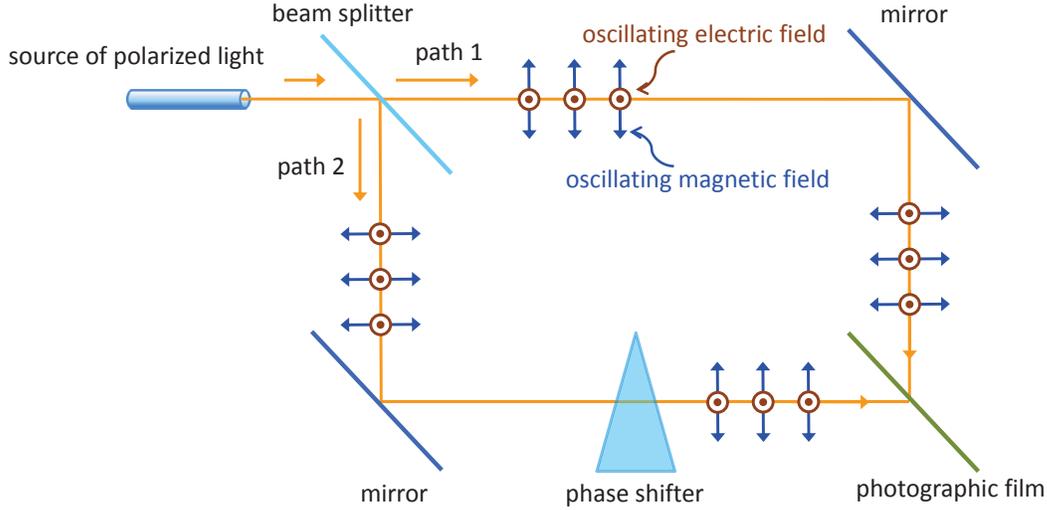}
}
\end{center}
\vspace{-22mm}
\caption{Mach-Zehnder-like interferometor. 
If the photographic film is sensitive to electric field,
an interference pattern will be observed.
If it is sensitive to magnetic field,
path of each photon will be distinguished.
}
\label{fig:1}
\end{figure}

In an interferometer depicted in FIG. 1,
the split light beams emerge on the film.
In this case, the oscillating electric fields of the two-way light incident on the film 
are parallel.
If we use a detector which is sensitive to electric field,
we cannot distinguish which-path of photons
and will observe interference pattern on the film.
On the other hand, 
the oscillating magnetic fields of the two-way light incident on the film 
are orthogonal.
Hence,
if we use an ideal detector which is sensitive to magnetic field polarization,
we can distinguish which-path of photons.
However, 
the wave nature and the particle nature of light
should not be simultaneously observed.
This complementary of the wave-particle natures 
is a mathematical consequence of the non-commutativity
or the uncertainty relation
of the electric and the magnetic fields
\begin{equation}
	[ \hat{E}_j ( \vect{x}, t ), \, \hat{B}_k ( \vect{y}, t ) ]
	=
	i \hbar \, \varepsilon_{jkl} \, \frac{\partial}{\partial x_l}
	\delta^3 ( \vect{x} - \vect{y} ).
	\label{commutator of EB}
\end{equation}
More detailed discussion on this issue will be published in another paper.

Here we summarize our discussion:
We derived the photon detection operator
by applying the indirect measurement scheme to quantum electrodynamics.
The photon detection probability depends on
both electric and magnetic dipole moments of the photon-detecting atom.
Their complementarity reflects the non-commutativity
of the electric and the magnetic fields.

Acknowledgements: I thank Prof. Y. Ohnuki for valuable discussions.
He should be regarded practically as a collaborator on this work.

\end{document}